# A Unified Grand Canonical Description of The Nonextensive Thermostatistics of The Quantum Gases: Fractal and Fractional Approach

**Fevzi Büyükkılıç[†] and Doğan Demirhan**

*Department of Physics, Faculty of Science, Ege University 35100 Bornova-İzmir, TURKEY*

**Abstract**

In this paper, the particles of quantum gases, that is, bosons and fermions are regarded as g-ons which obey fractional exclusion statistics. With this point of departure the thermostatistical relations concerning the Bose and Fermi systems are unified under the g-on formulation where a fractal approach is adopted. The fractal inspired entropy, the partition function, distribution function, the thermodynamics potential and the total number of g-ons have been found for a grand canonical g-on system. It is shown that from the g-on formulation; by a suitable choice of the parameters of the nonextensivity q, the parameter of the fractional exclusion statistics g, nonextensive Tsallis $(q \neq 1)$ as well as extensive (q=1) standard thermostatistical relations of the Bose and Fermi systems are recovered.

---

[†] Correspondence address: Department of Physics, Faculty of Science, Ege University, 35100 Bornova/İzmir/Turkey    e-mail:fevzi@fenfak.ege.edu.tr

# 1. Introduction

For thermodynamical systems which have high densities of particles at very low temperatures, the criteria of being classical $n\lambda^3 \ll 1$ where n is the density of the particles and $\lambda$ is the thermal wavelength is no longer fullfilled. That means for very small temperatures as well as for high densities the interaction in real systems are no longer negligible and quantum effects are exhibited. The influence of quantum effects can be very well suited by the Bose and Fermi gas models where the corresponding well known Bose-Einstein (BE) and Fermi-Dirac (FD) statistics have fundamental importance for the understanding of the physical systems. Indeed, a great number of phenomena have been explained within the framework of BE and FD quantum statistics. Are they, however, unique and adequate to enlighten all the physical properties of the nature? In this connection, where a fractal space is involved, the quantum systems are expected to violate the standard properties and the nonextensive thermostatistics [1-23] is required.

The second point to be raised is how one can reconcile the standard statistics with fractional statistics, e.g. para-statistics [24], intermediate statistics [25], fractional exclusion statistics[26-27], etc? Can BE and FD statistics be unified as well as to include many other statistics?

Therefore, by adopting the fractional exclusion statistics the generalization for quantum gases has been done within the context of nonextensive thermostatistics the so-called Tsallis thermostatistics and finally the thermostatistics for bosons and fermions are unified.

The proposed formalism basically relies upon two parameters, one is usually called the entropic index; $q \in R$, which takes care of the nonextensivity and it includes the standard extensive quantities as a special case when q is set to be unity and the other is the statistics parameter g which may have fractional values representing the ability of the occupancy of the single particle states by a g-on, with g=0 corresponding to usual bosons and g=1 fermions [28-30].

The outline of this paper is as follows; In sections 2 and 3, for the benefit of the readers the origin of the fractal inspired entropies and derivation of Tsallis entropy are stated. In section 4, the generalized entropy of g-on gas is obtained. In section 5, due to the great importance of the generalized mean occupation number for g-ons which includes the standard extensive BE and FD distribution as a special case where g is taken 0 for bosons and 1 for fermions, the generalized g-on distribution is derived following two different methods. In section 6, the partition function, the thermodynamics potential, the total number of a g-on system within the nonextensive formalism are obtained. In section 7, the basic equations of the thermodymamic potential and the total number of the g-on system are calculated as an application, and their relation to the Bose and the Fermi system are exhibited.

## 2. Measure, Spectrum of Fractal Dimensions and Fractal Inspired Entropies

Consider a non-interacting quantum gas of N g-ons which is not in thermodynamical equilibrium. Let the system be in heat and particle baths, in other words, the total energy E and the total number of g-ons N fluctuate, but the total energy and the total number of g-ons are conserved on the average, that is, a grand canonical ensemble is established. The accessible states of the g-on system is given by the set of the occupation numbers $\{n_1, n_2, ... n_k, ... n_K\}$ where $n_k$ denotes the number of particles in state k [31]. K is the number of partitions of N particles. When a measure is done on the set

$$M_q = \sum_{i}^{K} P_i^q . \qquad (1)$$

where $P_i$ is the probability of the i.th partition.

Let us introduce the definition $D_q$ as the spectrum of the fractal dimensions

$$D_q = \lim_{\delta \to 0} \frac{1}{q-1} \frac{\ln \Omega_q}{\ln \delta} \qquad (2)$$

or it can be written in an exponential form as

$$M_q = \Omega_q \approx \lim_{\delta \to 0} \delta^{(q-1)D(q)} . \qquad (3)$$

where $\delta$ is the resolution related to the measure. From the relation between the entropy and the spectrum of fractal dimension, one of the fractal inspired entropies which is known as Tsallis entropy can also be obtained. The relation between the entropy and the spectrum of fractal dimension is given as

$$S_q = -k_B \lim_{\delta \to 0} \ln \delta \; D_q \qquad (4)$$

where $k_B$ is the Boltzmann constant.

On the other hand, in view of Eq. (1) and Eq.(2), the spectrum of fractal dimension is obtained in terms of probabilities:

$$D_q = \frac{1}{q-1} \frac{\ln \Omega_q}{\ln \delta} = \frac{1}{\ln \delta} \frac{\ln \sum P_i^q}{q-1} \qquad (5)$$

or

$$D_q \approx \frac{1}{q-1} \frac{\sum_{i=1}^{K} P_i^q - 1}{\ln \delta} \qquad (6)$$

since

$$\ln M_q \approx \ln \Omega_q \approx \sum_i P^q{}_i - 1 \qquad (7)$$

Then when Eq.(6) is substituted into Eq.(4), Tsallis entropy is obtained:

$$S_q = -\frac{\sum_{i=1}^{K} P_i^q - 1}{q-1} \qquad (8)$$

where K can be regarded as the number of accessible states of the system. In the limit $q \to 1$ the well known Shannon entropy is recovered.

$$S_1 = -k_B \Sigma P_i \ln P_i \qquad (9)$$

### 3. Nonextensive Entropy of a G-on gas

The entropy of a quantum gas can be expressed in terms of the statistical weights. The relation between the statistical weight of a g-on gas which takes into account the statistics which it obeys and its thermodynamics is provided through statistical mechanics. From this point of view, entropy is a bridge between micro and macrophysics. Thus, the entropy of a g-on gas system has a very close relation with the statistics it obeys.

In this section, the fractal inspired entropy of a non-interacting g-on gas system, which is not in equilibrium has been obtained in terms of the occupation numbers.

The statistical weight of a g-on gas system which is not in equilibrium obeying the fractional exclusion statistics, is given by [27,28]

$$\Omega(g) = \prod_{k=1}^{K} \frac{[g_k + (n_k - 1)(1 - g)]!}{n_k! [g_k - gn_k - (1 - g)]!}. \tag{10}$$

There are exactly $\Omega(g)$ ways to distribute the indistinguishable $n_k$ g-ons over the $g_k$ states in a certain cell, where $g_k$ is the number of states in group k, i.e. the degeneracy, $n_k$ is the number of particles in these states (occupation numbers), g is a statistics parameter $0 \leq g \leq 1$, indicating the ability of a particle to occupy a single particle state, with g=0 corresponding to usual bosons and g=1 to fermions [36,37].

In order to find the nonextensive entropy of a g-on gas fractal dimension $D_q$ is needed. By using Eq. (10) and following the generalizing technique in Ref[15].

$$D(g,q) = \sum_{k=1}^{\infty} \frac{u_k^q - v_k^q - n_k^q}{(q-1) \ln \delta} \tag{11}$$

is obtained for the g-on system, where

$$u_k = g_k + (n_k - 1)(1 - g) \tag{12 a}$$

and

$$v_k = g_k - gn_k - (1 - g). \tag{12 b}$$

Eq.(11) can be rewritten as [15]

$$D(g,q) = \sum_{k=1}^{K} g_k \frac{u_k^q - v_k^q - n_k^q}{(q-1)\ln\delta} \qquad (13)$$

where for simplicity

$$u_k \to \frac{u_k}{g_k} \to 1 + n_k(1-g) \qquad (14\ a)$$

$$v_k \to \frac{v_k}{g_k} \to 1 - gn_k, \quad n_k \to \frac{n_k}{g_k} \qquad (14\ b)$$

are taken in place of the corresponding expressions.

Using Eq.(12), the fractal inspired entropy of g-on system is found:

$$S(g,q) = k_B \sum_{k=1}^{K} g_k \frac{[1 + n_k(1-g)]^q - [1 - gn_k]^q - n_k^q}{q-1} \qquad (15)$$

where Eq.(14 a), (14 b) are used. The moment order q, which appears in the entropy for the g-on system, thereafter will be called as entropic index; in case of $q \to 1$ limit Eq.(15) reads:

$$S(g,1) = k_B \sum_{k=1}^{K} g_k [1 + n_k(1-g)] \ln[1 + n_k(1-g)] - [1 - gn_k] \ln[1 - gn_k] - n_k \ln n_k. \qquad (16)$$

Eq.(16) may lead to the usual entropy of the Fermi system S(1,1) and of the Bose system S(0,1) for g=1, and g=0 respectively. Thus

$$S(g,1) = \mp k_B \sum_{k=1}^{K} g_k [\pm n_k \ln n_k \pm (1 \mp n_k) \ln(1 \mp n_k)] \qquad (17)$$

where upper signs should be taken for the Fermi systems and the lower signs for the Bose systems.

**4. Derivation of The Generalized Distribution Function of a G-on System**

We assume that the H-theorem is valid for the generalized g-on system whose entropy is S(g,q). Therefore, the H-theorem, can be used to find the distribution function of the g-on gas. In course of time, the entropy of the g-on system must incline towards a maximum in accordance with Boltzmann's H-theorem, i.e. equilibrium state. The problem is to find $\bar{n}_k$ such that the equation of entropy for the g-on gas has a maximum under the constraints [15]

$$N_q = \sum_{k}^{K} g_k n_k^q \tag{18a}$$

$$E_q = \sum g_k \in_k n_k^q \tag{18 b}$$

which means that the total particle number and the total energy of the g-on gas respectively are conserved. The constraints are imposed for a macrocanonical ensemble where the g-on gas of the total particle number and the total energy may fluctuate by conserving these quantities on the average. This method of Undetermined Lagrange multipliers take the constraints (18 a) and (18 b) into account.

For this aim, the expression

$$F(g,q) = S(g,q) - \alpha N_q - \beta E_q \tag{19}$$

is constructed. The values of α and β have been identified as $\beta = \frac{1}{T}$ $\alpha = -\beta\mu$ where T is temperature, μ is chemical potential and Boltzmann constant $k_B$ is set to unity. When Eqs.(15), (18 a) and (18 b) are substituted into Eq.(19), after a partial differentiation with respect to $n_k$ and equating $\delta F(g,q)$ to zero, the following equation is obtained

$$\sum_{k=1}^{K} \left\{ (1-g)[1+n_k(1-g)]^{q-1} + g(1-gn_k)^{q-1} - n_k^{q-1} - \alpha(q-1)n_k^{q-1} - \beta(q-1)\in_k n_k^{q-1} \right\} = 0. \tag{20}$$

where we assume $n_k, g_k \gg 1$ and Stirling formula is used. Since we have imposed two constraints by implementing two Lagrange multipliers, we may assume that here the variations $\delta n_k$ are mutually independent. Then each coefficient in Eq.(20) must vanish:

$$(1-g)[1+n_k(1-g)]^{q-1} + g(1-gn_k)^{q-1} - n_k^{q-1} - \alpha(q-1)n_k^{q-1} - \beta(q-1)\in_k n_k^{q-1} = 0 \tag{21}$$

thus

$$(1-g)[1+n_k(1-g)]^{q-1} + g(1-gn_k)^{q-1} = n_k^{q-1} + \alpha(q-1)n_k^{q-1} + \beta(q-1)\in_k n_k^{q-1} \tag{22}$$

Unfortunately, Eq.(22) does not allow us to find an analytical expression for $\bar{n}_k$. In order to have a solution for $\bar{n}_k$ now let us write the left hand side of Eq.(22) in a more compact form, i.e.

$$(1-g)[1+n_k(1-g)]^{q-1} + g(1-gn_k)^{q-1} \approx \sum_{r=0}^{q-1} C_{q-1}^r \left[(1-g)^{r+1} + (-1)^r g^{r+1}\right] n_k^r \approx$$
$$[1+(1-2g)n_k]^{q-1} \tag{23}$$

where $C_{q-1}^r$ is the usual combinatorial term. The term in the parenthesis on the left hand side of Eq.(23) is approximated as :

$$\left[(1-g)^{r+1} + (-1)^r g^{r+1}\right] \approx (1-2g)^r . \tag{24}$$

Then Eq.(22) reads

$$[1+(1-2g)n_k]^{q-1} = n_k^{q-1} + \alpha(q-1)n_k^{q-1} + \beta(q-1)\in_k n_k^{q-1}. \tag{25}$$

Now the most probable distribution of the $N_q$ g-ons over the single states is solved:

$$\bar{n}_k(g,q) = \frac{1}{[1-(1-q)\beta(\in_k -\mu)]^{\frac{1}{q-1}} + 2g - 1} \tag{26}$$

where $\alpha$ and $\beta$ are substituted into Eq.(26). $\bar{n}_k(g,q)$ can now be interpreted as the number of g-ons per energy level which unifies the form of the generalized quantum distributions for bosons and fermions. It is obvious that, for g=1 the generalized Fermi-Dirac $\bar{n}_k(1,q)$ and for g=0 the generalized Bose-Einstein distribution $\bar{n}(0,q)$ are recovered [15]. It should be remarked that, $\bar{n}_k(g,q)$ also includes the standard distribution functions for extensive systems that is $\bar{n}_k(1,1)$, $\bar{n}_k(0,1)$ $\bar{n}_k\left(\frac{1}{2},1\right)$ for quantum (fermi and bose) and classical systems, respectively. Eq.(26) may also lead to a generalized Planck distribution for µ=0 and g=0 [36-37]. Standard Planck distribution is recovered for µ=0, g=0 and q=1.

## 5. Generalized Grand Canonical Description of a G-on System

Because of the great importance of the generalized mean occupation number of a g-on gas we want to rederive it using the generalized grand canonical partition function. Along this line, we introduce for the first time a generalized partition function $Z(g,q)$ for grand canonical ensemble of a g-on gas within the factorization method which has beens developed by the authors of the present article[15];

$$Z(T,V,g,\mu,q) = \prod_{k=1}^{\infty} \left\{ \sum_{n_k=0}^{1} [1+(q-1)n_k x_k]^{\frac{1}{1-q}} + (1-g) \sum_{n_k=2}^{\infty} \left[ [1+(1-q)n_k x_k]^{\frac{1}{1-q}} \right] \right\} \quad (27)$$

where
$$x_k = \beta(\in_k - \mu). \quad (28)$$

is taken.

Eq.(27) can be regarded as a unified partition function of quantum gases. It is obvious that for g=0 the generalized partition function of the Bose gas $Z(T,V,0,\mu,q)$ and for g=1 the generalized partition function of the Fermi gas $Z(T,V,1,\mu,q)$ are included in Eq.(27). Following the method in ref.[15] the generalized distribution function of a g-on gas can be obtained by using Eq.(27), then, one ends up with the same given by Eq.(26).

## 6. Thermodynamics of the G-on Gas

In order to obtain the thermodynamics formulations of the g-on gas $\Phi(T,V,g,\mu,q)$ is required. The generalized thermodynamics potential of the g-on gas, however, is obtained by incorporating the partition function with the relation

$$\Phi(T,V,g,\mu,q) = -k_B T \ln Z(T,V,g,z,q) \quad (30)$$

where summation is done on all of the quantum states. By taking into account Eq.(27) one derives

$$\Phi(T,V,g,\mu,q) = -k_B T \sum_{k=1}^{K} \ln\left[1 + \frac{1}{\omega_k}\right] \quad (31)$$

where
$$\omega_k = [1-(1-q)x_k]^{\frac{1}{q-1}} + g - 1. \quad (32)$$

In $q \to 1$ limit, Eq.(31) may be compared with Eq.(4.2) of ref. [27]:

$$\Phi(T,V,g,\mu,1) = -k_B T \sum_{k=1}^{K} \ln\left[1 + \frac{1}{e^{x_k} + g - 1}\right]. \tag{33}$$

In the mean time, in view of Eq.(32), Eq.(26) can be rewritten in terms of $\omega_k$ as

$$\bar{n}_k(g,q) = \frac{1}{\omega_k + g}. \tag{34}$$

Since the generalized thermodynamics potential is

$$\Phi(T,V,g,\mu,q) = -pV \tag{35}$$

the pressure of a g-on gas is obtained:

$$p_q V = k_B T \sum_k \ln\left(1 + \frac{1}{\omega_k}\right). \tag{36}$$

While the total number of particles is found as

$$N_q = \sum_k \bar{n}_k(g,q) = \sum_k \frac{1}{\omega_k + g}. \tag{37}$$

The state of the gas is determined by Eq.(36) and Eq(37). The energy of the gas, however, is given by

$$E_q = \sum_k \bar{n}_k(g,q)\epsilon_k = \sum_k \frac{\epsilon_k}{\omega_k + g}. \tag{38}$$

where Eq.(34) is taken into account.

In order to calculate the thermodynamics potential, the total number of particles, internal energy, etc. of the g-on system, while including the correction due to the nonextensivity one may need the Taylor expansion of the distribution function of the g-on gas system. Therefore, the Taylor expansion of Eq.(26) is written:

$$\bar{n}_k(q,g) = \bar{n}_k(g,1) + \frac{1}{2}(q-1)x_k^2 \bar{n}_k^2(g,1) + O(q-1)^2 \tag{39}$$

where

$$\bar{n}_k(g,1) = \frac{1}{e^{x_k} + 2g - 1}. \tag{40}$$

It should be note that all terms in Eq.(39) are written in the exponential forms which mathematically are favourable. Apart from the first term, the rest of the terms can be interpreted as a correction due to nonextensivity of the entropy, each of which lead to zero in the limit $q \to 1$.

As a first concrete application of the generalized fractional exclusion statistics we want to calculate the properties of an ideal gas of nonrelativistic indistinguishable g-on system. Our aim is now to calculate the generalized grand canonical partition function, or more simply, its logarithm $Q(T,V,g,z,q)$, then the grand canonical thermodynamics potential according to Eq.(30) is found. In view of Eq.(31), $Q(T,V,g,z,q)$ is given as

$$Q(T,V,g,z,q) = \ln Z(T,V,g,z,q) = \sum_k \ln\left[1 + \frac{1}{\omega_k}\right]. \tag{41}$$

Let one of the g-on energies $\in_k$ are those of the energies of a free quantum mechanicle particle in a box of volume V. The chemical potential or fugacity z are not fixed, but the total number of g-ons are conserved on the average and z has to be determined from the Eq.(37). For a large volume the sum over all of the single-g-on states can be given as the single-g-on density states, where spin is not taken into account.

$$D(\in) = A \in^{\frac{1}{2}} \tag{42}$$

can be written, where

$$A = \frac{2\pi V}{h^3}(2m)^{3/2}. \tag{43}$$

Thus, the summation in Eq.(41) becomes

$$Q(T,V,g,z,q) = A \int_0^\infty \ln\left[1 + \frac{1}{\omega_k}\right] \in^{\frac{1}{2}} d\in \tag{44}$$

where Eq.(42) is taken into account. When the integration is performed by parts $Q(T,V,g,z,q)$ is reduced to:

$$Q(T,V,g,z,q) = \frac{2}{3}A\beta \int_0^\infty \epsilon^{3/2} \overline{n}(g,q) d\epsilon \tag{45}$$

After the substitution of the Eq.(39) into Eq.(45) and then rearranging,

$$Q(T,V,g,z,q) = \frac{2}{3}A\beta \left\{ \begin{array}{l} z\int_0^\infty \dfrac{\epsilon^{3/2}}{e^{\beta\epsilon}-(1-2g)z} d\epsilon + \dfrac{1}{2}(q-1)\beta^2 z \left[ \int_0^\infty \dfrac{\epsilon^{5/2} e^{-\beta\epsilon}}{\left(1-z(1-2g)e^{-\beta\epsilon}\right)^2} \right] \\ -2\mu \int_0^\infty \dfrac{\epsilon^{3/2} e^{-\beta\epsilon}}{\left(1-z(1-2g)e^{-\beta\epsilon}\right)^2} + \mu^2 \int_0^\infty \dfrac{\epsilon^{1/2} e^{-\beta\epsilon}}{\left(1-z(1-2g)e^{-\beta\epsilon}\right)^2} \end{array} \right\} \tag{46}$$

is obtained.

The integrals are performed using the related formulae of p 326 and p 330 of ref[38] since the necessary conditions are met. Thus one concludes that, for a nonextensive g-on system

$$Q(T,V,g,z,q) = \frac{2}{3}A\beta \left\{ zF_{5/2}(g,z) + \frac{1}{2}(q-1)\beta^2 z \left[ F_{7/2}(g,z) - 2\mu F_{5/2}(g,z) + \mu^2 F_{1/2}(g,z) \right] \right\} \tag{47}$$

where the definition

$$F_n(g,z) = \frac{\Gamma(n)}{z(1-2g)\beta^n} \sum_{k=1}^K \frac{[z(1-2g)]^k}{k^{n-1}} \tag{48}$$

is adopted in which $\Gamma(n)$ is the usual gamma function. Then the state of the nonextensive system is determined by the following equation:

$$pV = k_B T Q(T,V,g,z,q). \tag{49}$$

It is obvious that the value of $Q(T,V,g,z,q)$ for a bose system is obtainable by taking g=0 while for a fermi system by g=1. $\epsilon = 0$ plays a special role in the bose system since it is not taken into account by the integrals, therefore one has to explicitly account for the terms in the sums in eq.(37) and eq.(41) in the transition from summation to integration.

In the thermodynamics limit, that is in the limit of infinite volume with particle density held fixed, particles in the ground state, having no kinetic energy do not contribute to the pressure, then ,in terms of the thermal wavelength, the pressure of a Bose system become, as it is expected,

$$p = \frac{kT}{\lambda^3} \sum_{k=1}^{\infty} \frac{z^k}{k^{3/2}} \qquad (50)$$

where $\Gamma\left(\frac{5}{2}\right) = \frac{\sqrt{\pi}}{2^2} 3$ is taken.

In a similar manner, the thermodynamical properties of a Fermi system follows immediately from the logarithm of the grandpartition function of a g-on system, which is given by Eq.(47). In terms of the thermal wavelength, the pressure of the Fermi system is

$$p = g_s \frac{k_B T}{\lambda^3} \sum_{k=1}^{\infty} \frac{(-1)^{k-1} z^k}{k^{5/2}} \qquad (51)$$

where $g_s$ is due to spin.

### 7. The Total Number of The G-on Gas

For a large volume, in accordance with Eq.(37), the total number of g-ons, could be written as

$$N(T,V,g,z,q) = Az \int_0^{\infty} \frac{\epsilon^{1/2}}{e^{\beta\epsilon} - z(1-2g)} d\epsilon + \frac{1}{2}(q-1) A \beta^2 z \int_0^{\infty} \frac{(\epsilon-\mu)^2 \epsilon^{1/2} e^{-\beta\epsilon}}{(1 - z(1-2g) e^{-\beta\epsilon})^2} \qquad (52)$$

assuming that the single number g-on states can be calculated in terms of integrals. The performation of integrals in Eq.(52) using the related integrals in ref [38], p.326 and p 330 leads to:

$$N(T,V,g,z,q) = Az F_{3/2}(g,z) + \frac{1}{2}(q-1) A \beta^2 z \left[ F_{7/2}(g,z) - 2\mu F_{5/2}(g,z) + \mu^2 F_{3/2}(g,z) \right] \qquad (53)$$

where $F_n(g,z)$ is given by Eq.(48). This is the formula for the total number of a g-on system. If g=0 and g=1 is substituted in Eq.(53), the formula for the total number of a nonextensive Bose system $N(T,V,0,z,q)$ and for a nonextensive Fermi system $N(T,V,1,z,q)$ are recovered respectively.

For the total number of bosons in the extensive bose system, however, the parameter of statistic g=0 and the entropic index q =1 are taken in Eq.(53). Thus, for an extensive Bose system the total number of particles $N(T,V,0,z,1)$ including $\varepsilon = 0$ energy states is found to be :

$$N(T,V,0,z,1) = AzF_{3/2}(0,z) + \frac{z}{1-z} \tag{54}$$

where the last term $N_0 = \frac{z}{1-z}$ represents the contribution of the energy level $\epsilon = 0$ to the mean particle number and $F_{3/2}(0,z)$ is given by Eq.(48).

In a similar manner, for the total number of fermions of the extensive Fermi system the parameter of statistics g=1 and the entropic index q =1 are substituted in Eq.(53). Thus, for an extensive Fermi system, the total number of fermions $N(T,V,1,z,1)$ is derived.

The total number of fermions is find in terms of the thermal wavelength as

$$N(T,V,1,z,1) = g_s \frac{V}{\lambda^3} \sum_{k=1}^{\infty} \frac{(-1)^{k-1} z^k}{k^{1/2}} \tag{55}$$

where $\Gamma\left(\frac{3}{2}\right) = \frac{\sqrt{\pi}}{2}$ is taken, and this is an expected standard result [36,37].

## 8. Conclusions

In this study, a unified grand canonical description of the nonextensive thermostatistics of quantum systems have been achieved. This study has been initiated from the point of departure of bosons and fermions, which obey different statistics, could be regarded as a one-particle, simply, g-on which obey fractional exclusion statistics. Then using the fractional exclusion statistics where g-on number g is involved, and g might have fractional values as well, the thermostatistical relations which concerns quantum systems are unified, within the framework of nonextensive formalism which is introduced by C.Tsallis where entropic index q is involved. In this context for a grand canonical ensemble the fractal inspired entropy, the partition function, the distribution function, the thermodynamics potential, the total number of particles, the state of a g-on system are obtained. It is clearly shown that for g=0 and for g=1 the above mentioned quantities lead to corresponding quantities of a Bose and a Fermi system, respectively, while taking into account the nonextensivity of the systems involved. Furthermore, when the entropic index q is set to one, the corresponding quantities for Bose and Fermi extensive systems i.e. standard relations are recovered.

## Acknowledgment

Authors would like to thank Ege University Research Fund for their partial support under the Project Number 98 FEN 25.